**Correspondence: Reply to 'On the understanding of current-induced spin polarization of 3D topological insulators'**

C. H. Li[1*], O. M. J. van 't Erve[1], S. Rajput[2], L. Li[3], and B. T. Jonker[1]

[1]Materials Science and Technology Division, *Naval Research Laboratory*, Washington, DC 20375  USA.

[2]*Intel Corporation*, Hillsboro, OR 97124  USA.

[3] Department of Physics and Astronomy, *West Virginia University*, Morgantown, WV  26506 USA.

* Email: connie.li@nrl.navy.mil

We reported the first spin potentiometric measurement to electrically detect spin polarization arising from spin-momentum locking in topological insulator (TI) surface states using ferromagnet/tunnel barrier contacts [1]. This method has been adopted to measure the current generated spin in other TI systems [2-10], albeit with conflicting signs of the measured spin voltage [1,2,4,6-10]. Tian *et al.* wish to use their model as presented in Ref. [4] to determine the sign of the induced spin polarization, and thereby determine whether the claims of various groups to have sampled the topologically protected surface states in bulk TIs are correct. The central point of our Reply is that the model as presented is incapable of doing so because it fails to include separate physical contributions which independently effect the sign of the spin polarization measured.

In our recent paper [9], as stated therein, we adopted the generic framework of Tian *et al.* [4] which postulates parallel spin dependent potentials in the TI channel, to be consistent with literature discussions. The issue raised by Tian *et al.* is the specific ordering of the spin-up and spin-down electrochemical potentials within their original model. We had placed the spin-up (majority) electrochemical potential level further from zero due to its greater magnitude [9], which may be inconsistent with their original model. We subsequently realized that this model neglects critical spin dependent parameters such as interface and channel resistances, and believe that this model is too simplistic to accurately account for the real experimental systems under study. Specifically, the potential profiles include vertical drops at the contact interface to the TI channel, which would indicate a current discontinuity (i.e., an infinite current), and therefore is fundamentally unphysical. It is this inadequacy which has resulted in apparent contradictory results from different groups. We have developed a more realistic model that includes experimental parameters such as interface and spin-dependent channel resistances. We find that such considerations can cause a crossing of the spin-up and spin-down voltage levels within the TI channel, which can lead to measured spin voltages of either sign depending upon the magnitude of the spin dependent resistances at interfaces and in the channel [11]. Our model potentially reconciles the inconsistent signs of spin voltages reported in the literature.

Any practical system will have contacts for experimental access, and these interfaces are crucial to electrical transport when current is converted from one type of charge carrier to another. The interface resistances at the current injecting contacts may be nonlinear or larger than that of the TI channel, perhaps due to oxidation of the TI surface or a blanket layer of tunnel barrier material such as $Al_2O_3$ often deposited for capping purposes and/or to simplify fabrication [1,2,4-10]. This creates interface resistances of varying degree that can fundamentally change the carrier conversion at the interface. Fig. 1a shows a typical *I-V* curve taken between the current injecting $Au/Al_2O_3/Bi_2Se_3$ contacts, showing a non-negligible and nonlinear interface resistance where a voltage drop can be supported.

Fig. 1b shows a schematic resistor circuit model for spin-up and spin-down electrons traveling in two independent channels [12-14]. Each component of the circuit, including the contacts and interfaces, is modeled as a resistor. We have used a similar approach to model spin filtering in graphene magnetic tunnel junctions [15]. As electrons travel from the right gold electrode to the left, several resistances are encountered: the right Au electrode $R_{Au,R}$, the right Au/Al$_2$O$_3$/TI interface $R_{int,R}$, TI channel resistance $R_{TI}$, the left TI/Al$_2$O$_3$/Au interface $R_{int,L}$, and the left Au electrode $R_{Au,L}$.

For electrons traveling from the right Au electrode, the resistance of the Au electrode is low for both spin-up and spin-down electrons. However, the interface resistance for spin-up and spin-down electrons entering into the TI channel may be different depending upon their alignment with the states in the TI. A steady state electron current flowing through the TI surface states from right to left creates a spontaneous spin-up orientation on the top TI surface due to spin momentum locking. Hence for spin-up electrons, this interface resistance will be lower since they align with those in the TI surface states [16]. The opposite is true for spin-down electrons – the interface resistance will be higher due to their antiparallel alignment, i.e., ***$R_{int,R\uparrow}$<$R_{int,R\downarrow}$***. Within the TI channel, the resistance for the spin-up electrons will be significantly lower due to the available spin-up states arising from the left flowing electron current, and higher for the spin-down electrons. Finally, as these electrons enter into the left Au electrode, the interface resistance will be similar for both spins since there are equal number of spin-up and spin-down states in the Au, i.e., the interface resistance here will be spin *in*dependent, i.e., ***$R_{int,L\uparrow}$=$R_{int,L\downarrow}$***.

Given that the overall voltage drop for both the spin-up and down channels must be the same across the left and right Au electrodes, and that the spin-up channel is clearly a lower resistance channel, the current flowing through the spin-up channel (***$I_\uparrow$***) will be greater than that for the spin-down channel (***$I_\downarrow$***), or ***$I_\uparrow$>$I_\downarrow$***.

At the left TI/Al$_2$O$_3$/Au interface that is spin *in*dependent, or ***$R_{int,L\uparrow}$=$R_{int,L\downarrow}$***, the voltage drop at the interfaces is greater for the spin-up (***$V_{int,L\uparrow}$***) than the spin-down (***$V_{int,L\downarrow}$***) channel because $I_↑>I_↓$, as depicted by the steeper slope for the spin-up (blue) lines within the left TI/ Al$_2$O$_3$/Au interface region in Fig. 2a.

Conversely, at the right Au/Al$_2$O$_3$/TI interface that is spin-dependent, the voltage drop for the spin-up and spin-down electrons ($V_{int,R↑}$ and $V_{int,R↓}$, respectively) can have two different outcomes: $V_{int,R↑}<V_{int,R↓}$ or $V_{int,R↑}>V_{int,R↓}$, as shown in Fig. 2a and b respectively, depending on the relative magnitudes of the currents through the spin-up and spin-down channels ($I_↑$, $I_↓$), compared to that of the spin dependent resistances at the interface ($R_{int,R↑}$ and $R_{int,R↓}$). The case for $V_{int,R↑}<V_{int,R↓}$, due for example to $I_↑≥I_↓$ and $R_{int,R↑}<<R_{int,R↓}$, is depicted by the steeper slope for spin-down (red) bands at the right Au/Al$_2$O$_3$/TI interface in Fig. 2a. Connecting the end points of the voltage profiles for spin-up and spin-down channels yields a profile where the spin-down band is consistently above the spin-up band throughout the channel.

However, in the case that $V_{int,R↑}>V_{int,R↓}$, due to for example $I_↑>>I_↓$ and $R_{int,↑}≤R_{int,↓}$, connecting the end points of the voltage profiles for spin-up and spin-down channels creates a crossing (Fig. 2b). It is important to recognize that this crossing does not imply a change in sign of the spin polarization in the channel or direction of the charge flow, but will result in a reversal of the measured spin voltage loop, as discussed below. Here charge carrier conversion at the Au/TI interfaces create boundary conditions to ensure the spin and current continuity across the interface that must be met, just like carrier conversion at any ferromagnet/nonmagnet interfaces [12-14]. This results in the splitting of the spin-up and spin-down levels near the interface in the TI. However, the spin coherence length in the TI is very large and typically greater than the TI channel length. Hence the equilibrium conditions that would be expected for an isolated TI are not reached, which allows for a current to exist where the chemical potentials overlap. Thus different voltage profiles and relative energies of the spin-up and spin-down levels can be created depending on the relative magnitudes of the spin-dependent resistances at interfaces and within the channel. Figures 2a and 2b clearly illustrate that opposite signs can indeed be measured for the spin voltage. A detailed treatment is provided in Ref. 11.

The probing of the voltage profiles for the spin-up and spin-down electrons by a ferromagnet/tunnel barrier contact via potentiometric measurement is the same as described originally [9]. This yields hysteresis loops of opposite signs as shown in Figs. 2c and d for the profiles in Figs. 2a and b, respectively. The sign in Fig. 2c is consistent with the observations in Refs. 4,6,8, and that of Fig. 2d is consistent with the observations reported in Refs. 1,8-10.

As noted above, the resistances at the left TI/$Al_2O_3$/Au and right Au/$Al_2O_3$/TI interfaces are not symmetric, because they are spin dependent when entering the TI channel, and spin-*in*dependent when entering the Au electrode. The metal/TI current injecting contacts are typically non-ohmic and/or rectifying, due to TI surface oxidation or inclusion of a tunnel barrier [1,2,4-10]. This results in a junction where the magnitudes of these two interface resistances can vary depending on the current direction. This is depicted by the larger voltage drop at the higher resistance interface (entering the TI channel) in Figs. 2a,b. This asymmetry leads to a larger splitting between the spin-up and spin-down voltage levels at the higher resistance interface, pushing the crossing towards the opposite end of the TI channel (Figs. 2b). Hence, the spin signal probed at points along the TI channel may indeed be of the same sign, although a narrow detector contact placed very close to the opposite end of the TI channel (entirely on the opposing side of the crossing) would detect the opposite sign.

Regarding the work of Refs 9-12 on current generated spin in InAs cited in Tian et al.'s Correspondence, we note that different physical quantities were measured: Johnson/Hammar presented their results in voltage [Tian's Ref. 9,10] whereas Park et al. have shown resistance [Tian's Ref. 11,12], and hence these results are not directly comparable in terms of signs. We further note that these papers provide inadequate documentation of current and/or magnetic field direction and/or of the polarities of voltage probes to enable a reader to accurately establish a sign convention for their current generated spin polarization. However, they do report a value for the Rashba spin-orbit coupling coefficient, α, which is more fundamental and directly addresses the spin ordering of the states. The values for alpha reported in the papers by Johnson/Hammar and by Park et al. are positive. We specifically addressed this in our Nature Communications 2016 paper [9] by including a short discussion on page 7, first column, first full paragraph.

Regarding the experiments on $(Bi,Sb)_2Te_3$ by Lee et al. [5], the sign convention used there is ambiguous, because the magnetic field direction was indicated by double arrows, with no positive direction identified. We cited it as the same sign as our work based on statements made by one of the authors at a conference presentation.

In summary, the original model [4] which we adapted [9] is too simplistic and unphysical, and any discussion of the precise placement of the spin potentials is irrelevant to the determination of the sign and origin of the measured spin voltage. Our revised model remedies this by taking into account common experimental parameters such as spin-dependent channel and interface resistances, which can have a critical impact on the profiles and relative ordering of the spin dependent potentials, and the resultant spin voltage measured. Our conclusion in [9] that the spin signal we measured originates from $Bi_2Se_3$ Dirac surface states is sound -- it is independently supported by the different temperature dependences we reported for $Bi_2Se_3$ and InAs, and the self-cancellation of contributions from the spin-split 2DEG states which potentially coexist on the $Bi_2Se_3$ surface [17]. The opposite sign of the measured spin voltage we reported for $Bi_2Se_3$ relative to the InAs reference provides corroborative evidence, but we now realize is insufficient on its own unless the analysis includes the details of our new model [11] briefly summarized here. In addition, the InAs(001) surface 2DEG provides an excellent reference for the sign of the spin voltage in such an experiment, because the Rashba coefficient is well established, no additional states are present, and ultra-low resistance ohmic contacts can be obtained. Finally, our findings potentially reconcile the inconsistencies reported in the literature, and underscore the importance of recognizing these contributions in the interpretation of such data.

**Contributions**

All authors contributed to the discussion, analysis, and writing of the manuscript.

**Competing interests**

The authors declare no competing financial interests.

**Data availability**

The data sets generated during and/or analyzed during the current study are available from the corresponding author on reasonable request.

**FIGURE CAPTION**

**Fig. 1** (a) Typical I-V curve taken at 8 K between current injecting Au/Al$_2$O$_3$/Bi$_2$Se$_3$ contacts, showing a nonlinear behavior. (b) Schematic of a resistor circuit model for spin-up and spin-down electrons traveling in two independent channels from the right to the left electrode, where each component of the circuit from the contacts to interfaces are modeled as a resistor.

**Fig. 2** Voltage profiles for the spin-up (blue) and spin-down (red) electrons for a left flowing current, for the case (a) $V_{int,R\uparrow} < V_{int,R\downarrow}$ (due to for example $I_\uparrow \geq I_\downarrow$, $R_{int,R\uparrow} \ll R_{int,R\downarrow}$), and (b) $V_{int,R\uparrow} > V_{int,R\downarrow}$ (due to for example $I_\uparrow \gg I_\downarrow$, $R_{int,R\uparrow} \leq R_{int,R\downarrow}$). Predicted lineshape for the spin voltage measured by a ferromagnet/tunnel barrier detector contact for the voltage profiles in (a) and (b) are shown in (c) and (d), respectively.

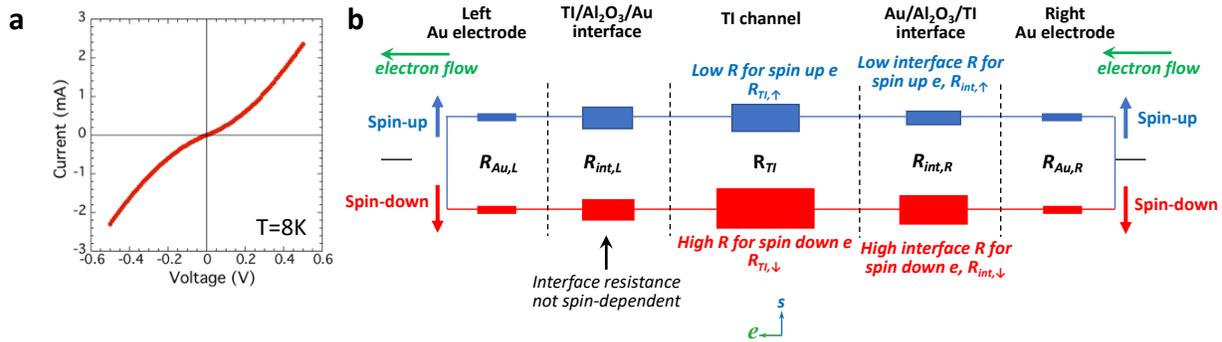

Li et al., Fig. 1

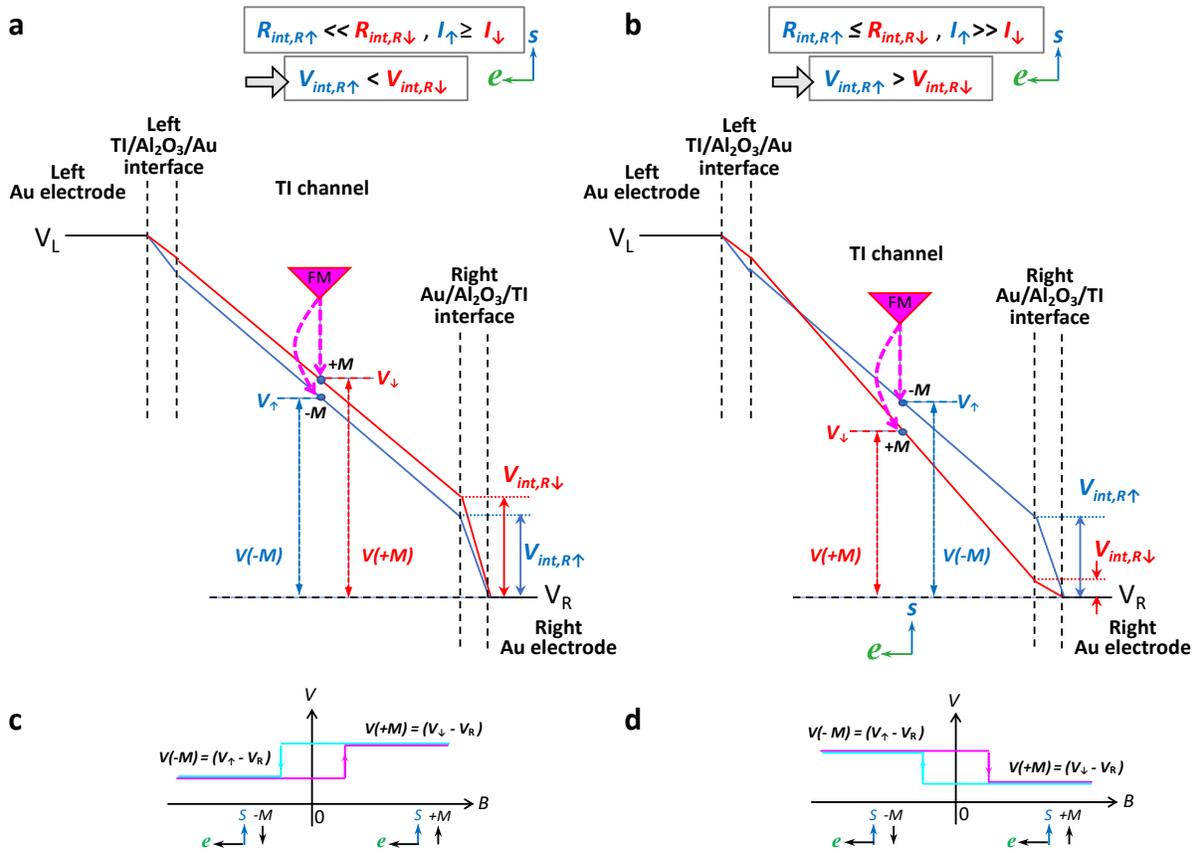

Li et al., Fig. 2

11